\documentclass{mem}
\usepackage{natbib}\usepackage{txfonts}\usepackage{balance}
\usepackage{graphicx}
\usepackage[a4paper,breaklinks,dvipdfm]{hyperref}
\usepackage{natbib}

\def\arcmin{\hbox{$^\prime$}}
\def\arcsec{\hbox{$^{\prime\prime}$}}

\def\apjl{ApJ~}

\def\aap{A\&A~}

\def\apjs{ApJS~}

\def\swi{Neil Gehrels Swift Observatory}

\begin{document}

\title{The SVOM mission, a pathfinder for THESEUS}
   \subtitle{}

\author{
B. Cordier\inst{1}
\and
D. \,G\"{o}tz\inst{1}
\and
C. Motch\inst{2}, on behalf of the SVOM collaboration\inst{3}
}

\institute{
CEA Saclay - Irfu/D\'epartement d'Astrophysique,
Orme des Merisiers, B\^at. 709,
F-91191 Gif-sur-Yvette, France
\email{bcordier@cea.fr}
\and
Universit\'e de Strasbourg, CNRS, Observatoire astronomique de Strasbourg, UMR 7550, F-67000 Strasbourg, France
\and
CAS, CNSA, NAOC Beijing, IHEP Beijing, NSSC Beijing, SECM Shanghai, CNES, LAM Marseille, IRAP Toulouse, GEPI Meudon, IAP Paris, LAL Orsay, CPPM Marseille, APC Paris, LUPM Montpellier, University of Leicester, MPE Garching, UNAM Mexico
}

\authorrunning{B. Cordier et al. }

\abstract{
The Sino-French space mission SVOM (Space-based multi-band astronomical Variable Objects Monitor) is mainly designed to detect and localize Gamma-Ray Burst events (GRBs). The satellite, to be launched late 2021, embarks a set of gamma-ray, X-ray and optical imagers. Thanks to its pointing strategy, quick slew capability and fast data connection to earth, ground based observations with large telescopes will allow us to measure redshifts for an unprecedented sample of GRBs. 
We discuss here the overall science goals of the SVOM mission in the framework of the multi-wavelength and multi-messenger panorama of the next decade. Finally we show how some developments of the SVOM mission will be helpful for the THESEUS project.

}
\maketitle

\section{Introduction}

The SVOM\footnote{SVOM stands for Space-based multi-band astronomical Variable Objects Monitor} project is based on a bilateral collaboration between France (CNES) and China (CAS, CNSA), with additional contributions from the University of Leicester, from the Max Planck Institut f\"ur Extraterrestische Physik and from the Universidad Nacional Autonoma de Mexico. The project is currently in Phase C with a scheduled launch late 2021. Nominal operations will last 3 years with a possible mission extension of 2 years. 

The main science goal, dubbed as the "core program" is Gamma-ray burst (GRBs) physics. This includes the GRB-Supernova connection, the nature of the central engine, the identification of the short GRB progenitors and the nature of the burst radiation processes. SVOM is designed to enhance the identification rate of high redshift GRBs. As such the core program is strongly related to cosmology with studies of the chemical enrichment of the interstellar medium, of the first stars and the of re-ionization of the Universe. SVOM will also carry out a general program and a target of opportunity (ToO) program that will both address non-GRB science (see section \ref{svomprograms}). 

All scientific prospects of the SVOM mission are described in \citet{wei2016}. In addition to advancing GRB physics, in particular at high redshift, SVOM will also be a major facility in the era of multi-messenger astrophysics and will help to identify gravitational wave counterparts and their relations to GRBs. SVOM will also bring crucial information on the X-ray transients possibly associated with cosmic neutrinos and very high energy photons. 

\section{SVOM instrumentation and system}

The SVOM platform embarks four instruments aiming at optimizing the rate of GRB identification. SVOM is designed as a multi-wavelength observatory with the capability to slew rapidly the narrow field X-ray and optical instruments in the direction of the detected burst while at the same time sending to ground based optical and infrared telescopes the preliminary position of the burst, later refined on board by the narrow field instrument.  In this respect, the SVOM project is similar to the very successful \swi. However, the soft response of the ECLAIRs wide field camera and fast links to ground based infrared telescopes should facilitate the identification of high redshift events.

\subsection{Space instrumentation}

\begin{figure*}
\center
\includegraphics[width=0.8\textwidth]{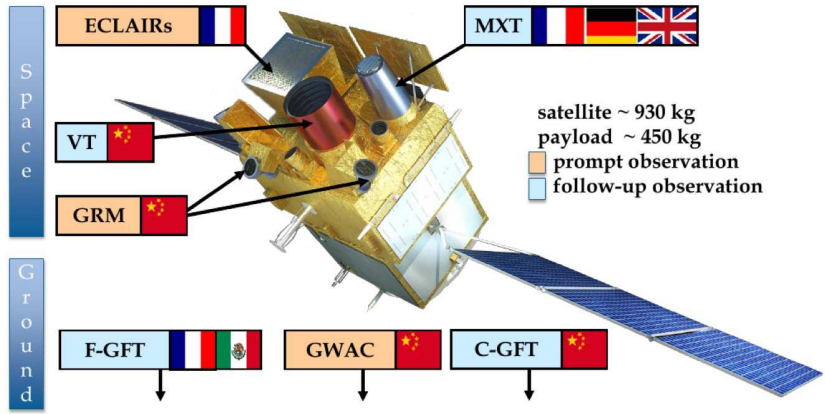}

{\caption{The instrumentation on board SVOM and the ground-based system.\label{instrumentation}}}
\end{figure*}

Fig. \ref{instrumentation} depicts the various instruments attached to the SVOM space segment.

ECLAIRs is a coded mask camera covering a wide field of view of 2\,sr at zero sensitivity in the energy range from 4 to 150\,keV \citep{godet2014}. The detection plane is made of 6400 CdTe pixels offering an effective area of 1024\,cm$^{2}$. The camera can localize X-ray and soft $\gamma$-ray transients with an accuracy better than 12\arcmin\ for 90\% of the sources at the detection limit. The on board detection algorithm will trigger and localize about 60 GRBs per year.

The Gamma-Ray Monitor (GRM) is made of 3 Gamma-Ray Detectors (GRDs). The field of view of this instrument is somewhat wider than that of ECLAIRs, each module covering 2 sr. Although the instrument only provides rough positions on the sky, it usefully extends the measured energy range up to 5,000 keV. A total of about 90 GRBs per year should be detected by the GRM \citep{bernardini17}. 

Two narrow fields instruments complement this observatory. First the Microchannel X-ray telescope (MXT) based on micro-pores optics in a Lobster Eye configuration offers a 27 cm$^{2}$ (1\,keV) effective area in its central spot \citep{mxt} and is sensitive in the 0.2-10 keV energy range.  Its field of view is of 57\,$\times$\,57\,arc min$^{2}$. The MXT will detect 90\% of the GRB X-ray afterglows and will be able to localize them of with an accuracy of 13\arcsec\ in 50\% of the cases. With a detection sensitivity of $\sim$ 10$^{-12}$\,erg\,cm$^{2}$\,s$^{-1}$ reachable in a 10\,ks, the MXT will be able to follow the evolution of the afterglow emission during at time interval of typically 24\,h after the initial trigger. 

The last narrow field instrument is the Visible Telescope (VT). It is a 40\,cm diameter Ritchey-Chretien telescope with a field of view of 26\,$\times$\,26 arc min well suited to the size of the error boxes delivered by ECLAIRs. With a V band sensitivity of V\,=\,22.5 in 300\,s, the VT will be able to localize with an exquisite accuracy of less than 1\arcsec\ GRBs up to $z$\,=\,6.5. 

\subsection{The ground-based system}

Three optical/infrared instruments make up the ground-based SVOM system. The first unit of the Ground-based Wide Angle Camera (GWAC), currently located Xinglong (China) is already operational. Another unit will be installed at CTIO (Chile). Each unit consists of 36 cameras covering 5400\,deg$^{2}$ in total in the wavelength range 500-850\,nm. GWAC can reach V\,=\,15 in 10\,s. 

Two robotic 1-m class telescopes put the final touch to the SVOM system. The Chinese Ground Follow-up Telescope (C-GFT) located at the Xinglong observatory
has a field of view of 21\,$\times$\,21\,arc min$^{2}$ in the 400-950\,nm wavelength range. The French Ground Follow-up Telescope (F-GFT) will be installed in San Pedro Martir (Mexico). It offers a field of view of 26\,$\times$\,26\,arc min$^{2}$ and operates simultaneously in 3 bands from 400 to 1700\,nm.

\subsection{The SVOM alert processing}

The SVOM alert processing is somewhat similar to that of Swift with the difference that SVOM does not use the NASA Tracking and Data Relay Satellite System. As soon as the wide field ECLAIRs camera detects a transient event on board, SVOM triggers a slew of the narrow field instruments toward the preliminary transient position in less than 5 minutes. X-ray and optical telescopes then acquire the field and detect an excess emission (afterglow) in $\sim$90\% of the cases. As soon as available, SVOM will send down to earth wide field and narrow field positions (in addition to some crude photon information) through a network of $\sim$\,45 VHF receivers distributed in longitude and covering a range of latitude of $\pm$\,30\,deg around the equator. Alerts are sent automatically to the SVOM ground based telescopes. After human qualification by a "Burst Advocate" alerts are distributed to the international community through the Gamma-ray Coordination Network and using Virtual Observatory events.

\subsection{The SVOM attitude law}

During the first three years of the nominal mission, SVOM will mostly point toward a roughly anti-solar direction, applying constraints on the Sun angle and avoiding the presence of bright X-ray sources such as Sco X-1 in the field of view of the coded mask ECLAIRs instrument so as to keep background as low as possible. 
 With this particular "B1" attitude law, GRBs will be detected on the night side allowing for fast ground-based follow-up. For instance, the median r-band AB magnitude of short GRBs afterglows is 23 at 7 hours for short GRBs, an order of magnitude lower flux than from long GRB afterglows. 

However, during about 15\% of the nominal mission time, SVOM will be allowed to point away from the "B1" attitude law in order to quickly follow-up ToO alerts from ground based optical or space borne X-ray telescopes e.g. supernovae, tidal disruptive events, but also events triggered by gravitational wave, VHE photons and neutrino detectors. The fraction of time devoted to external ToOs may rise to 40\% after three years of operation.

\begin{figure*}
\center
\includegraphics[width=0.8\textwidth]{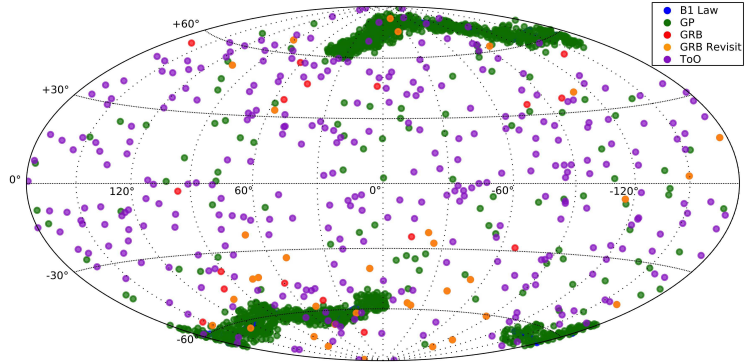}
{\caption{Simulation of the distribution in galactic coordinates of the \textit{SVOM} pointing direction of the narrow field of view MXT and VT instrument after a one year long mission time. \label{attitudelaw}}}
\end{figure*}

\subsection{SVOM observing programs}\label{svomprograms}

\begin{figure}
\includegraphics[width=\columnwidth]{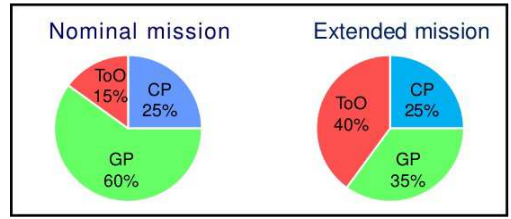}
{\caption{Allocated fraction of observing time to each SVOM scientific programs.\label{allocated}}}
\end{figure}

The SVOM science management plan includes three distinct observing programs. The relative time devoted to each program is bound to evolve during the mission (see Fig.\ref{allocated}).

The Core Program (CP) is dedicated to GRB studies. Alerts triggered by ECLAIRs will be monitored during several days after the initial event by the SVOM space-borne and ground-based instruments. The GRB data products (position, light curve, pre-computed spectra, etc...) will be made public immediately after validation. 

SVOM is also meant to be an open observatory and will carry out a General Program (GP). GP observations will be awarded by a TAC (a SVOM co-I needs to be part of your proposal). Although most of the GP targets are expected to be within the reach of the "B1" attitude law, during the nominal mission, about 10\% of the GP time will be offered on targets away from the attitude law, such as low Galactic latitude sources. This fraction may increase up to 50\% during the extended mission.

The Target of Opportunity program consists in triggering from the ground the follow- up of alerts sent by SVOM itself and by other observatories. For instance, SVOM may detect X-ray transients on board by the ECLAIRs trigger module or may identify them on the ground thanks to an off-line processing of the instrument data. It is initially foreseen to observe with SVOM a maximum of one ToO per day with science focused on time domain astrophysics. For multi-messenger alerts and exceptional astrophysical events, SVOM can perform one ToO every month with a delay between alert and observation reduced to 12 h. The ToO program time will significantly increase during the extended mission. 

Fig. \ref{attitudelaw} shows a simulation of the SVOM sky coverage for the different observing programs after one year.

\section{SVOM in the multi-wavelength and multi-messenger era}

New large sky area time domain oriented instruments, like the LSST in the optical or the SKA at radio wavelengths, will deliver an unprecedented number of well localized ToOs that may be worth following up with the space-borne and ground-based SVOM instruments.  One of the main issue will be then to sort out the alerts that may be most adapted to the instrumental capabilities and science goals of SVOM. 

Gravitational wave detectors (advanced LIGO, VIRGO, LIGO-India, KAGRA, etc.) will provide more alerts than now with improved localisation accuracies. Neutrino detectors IceCube-Gen2, KM3NeT (extension of ANTARES), etc. will also trigger a significant number of alerts. 

In this respect, the SVOM/MXT grasp appears to be well adapted to the large error boxes delivered by gravitational wave and neutrino detectors. The one square degree area covered in a single MXT observation largely compensates for the higher sensitivity provided by the narrower field of view XRT on board the \swi. The MXT is thus able to survey rapidly large sky areas at a sensitivity roughly similar to that of the XRT.

\subsection{Can SVOM help to unveil the origin of cosmic neutrinos ?}

IceCube results indicate that cosmic neutrinos are not associated with any particular class of objects, particularly GRBs  \citep{gonzalez2017,icecube2017}.  AUGER confirms the lack of association \citep{nellen2017}. 

Detecting a high-energy neutrino signal in coincidence with a GRBs would be a direct proof of the existence of a hadronic component in the jets. Proton interactions with the fireball photons can produce a burst of neutrinos with energies of $\sim$ PeVs \citep{kumar2015}. In the absence of simultaneous GRB / neutrino detection so far, one can conclude that the standard GRB population is not a major contributor to the diffuse high energy neutrino flux.  

However, low-luminosity GRBs associated with hypernovae, largely missed by currently operating $\gamma$-ray satellites (but more easily seen by SVOM due to the softer response of the wide field ECLAIRs camera) may contribute significantly to the high energy neutrino diffuse detections \citep{wang2007} and may provide an opportunity to identify the astrophysical source of some high energy neutrinos.

Inversely, both the space-borne and ground-based SVOM instruments will be able to follow-up alerts triggered by IceCube \citep{icecubealert}. 

\subsection{Finding the origin of VHE gamma-ray transient sources}

The High Altitude Water Cherenkov Experiment (HAWC; Mexico) maps the sky at gamma ray energies above 1 TeV. HAWC has the capability to detect and localize transient sources with an accuracy of 1 deg or less and issue rapid notifications of flares \citep{2017ApJ...843..116A}. CTA in WF/Survey mode will also detect transients. SVOM may detect with ECLAIRs events coincident with VHE transients, but it is also anticipated that both the space-borne and ground-based telescopes respond to such transient alerts.  The size of the error boxes delivered by the very high energy laboratories will be small enough to initiate follow-up observations with the MXT in one single MXT tile. 

\subsection{SVOM and the gravitational waves}

On August 17$^{th}$ 2017 a gravitational wave associated to the merger of a binary neutron star system has been detected for the first time by the aLIGO/Virgo
interferometers \citep{abbott17a}. A simultaeous $\gamma$-ray signal was detected by the \textit{Fermi} and \textit{INTEGRAL} satellite \citep{goldstein,savchenko} marking the beginning of multi-messenger astrophysics. The realtivey good location accuracy of this event ($\sim$ 30 deg$^{2}$) allowed for the detection and identification of a bright optical counterpart (AT2017gfo) in the galaxy NGC 4993 which turned out to be conisitent with the prediction of a ``kilonova'', i.e. a source powered by the radioactive decay of heavy nuclei produced through rapid neutron capture \citep[see][and references therein]{pian}. 

Short GRBs and their afterglows are the main counterpart contenders of merger events involving a neutron star \citep[see e.g.][]{metzger2012}. In this case, the GRB jet created by accretion of matter onto the newly formed BH is beamed towards the earth \citep[see e.g. ][and references therein]{nakar2007,berger2014}. Radioactive decay of the unstable nuclei created in the neutron-rich ejecta can power a strong near infrared emission \citep[see][for a recent review]{metzger2017}. Finally, the interaction of the ejecta with the surrounding interstellar material, may produce a surge of radio emission. 

The short GRB emission typically lasts less than 2\,s and can be detected by ECLAIRs and by the GRM. X-ray afterglow emission was observed for 89\% of the short GRBs detected by the \swi~ and fades below detection after about $\sim$ 1,000\,s at the sensitivity of the SWIFT XRT. Prompt optical emission was seen at about 21-26 mag between 0.4 and 30\,h after the GRB \citep{berger2014}. Since the SVOM platform slews to the ECLAIRs position in less than 5\,min, the MXT and VT instruments have a high probability to detect these afterglows. Again, the SVOM attitude law, pointing towards the anti-sun direction will allow quick optical follow-up from the ground.  

Even in the absence of X-ray detection, the near-infrared sensitive GFT will be able to survey the $\sim$ ECLAIRs error box searching for kilonovae candidates whose emission peaks a few days after the short GRB. On the other hand, the surge of radio emission spreads over much longer time scales, months to years.   

Importantly only a relatively small fraction of the mergers involving a neutron star will produce a short GRB and thus trigger a SVOM slew. A jetted component may not be created in the merging process, or the direction of the jet responsible for the $\gamma$ and X-ray emission may not sweep the earth. However, it has been proposed that the variable extended X-ray emission exhibited by some short GRBs could be due to the creation of a post-merger rapidly rotating magnetar neutron star \citep{zhang2013}. In this case, X-ray emission is expected to be less beamed than that of the GRB and possibly last thousands of seconds, opening the possibility to link GRB-less X-ray transients to gravitational waves. 

The LIGO alert system is able to broadcast GW positions about 30\,minutes after the event\footnote{see http://www.ligo.org/science/Publication-S6EMFollowUpMethods/}. Most of the latency time is due to human verification and could be shortened were efficient false event filtering tool be implemented \citep{ligotrigger2012}. 

Taking advantage of the wide field of view of the X-ray telescope of SVOM, the ToO SVOM program will implement a fast X-ray and optical survey of the error boxes delivered by the gravitational wave detectors using an optimal strategy so as to target the best candidate galaxies. With the advent of the advanced LIGO, advanced Virgo and LIGO-India in the near future, source localization by triangulation will provide error boxes as small as 5\,deg$^{2}$ and 20\,deg$^{2}$ in 20 and 50\% of the cases in 2022 \citep{abbott2016}. Such relatively small sky areas can in principle be quickly surveyed by the MXT on board SVOM.  

\section{SVOM as a pathfinder for THESEUS} 
The development of the SVOM mission will enable the development of concepts and objects that will be reused by the THESEUS mission. 
With a SVOM launch date planned for 2021, all these concepts /objects will have reached a great maturity at the time of the THESEUS mission adoption.

\subsection{Micro-Pores Optic (MPO) validation plan within the MXT project}
The MXT optics is a small FOV version of the SXI optics (see O'Brien, these proceeding). It is made by 5$\times$5 MPO plates of 1-2 mm thickness, 4 cm side, 
composed by 40 $\mu$m square pores.

In the framework of the SVOM project, during MXT phase B, the Photonis Company produced bread board of MPOs and a complete set of Structural and Thermal Model of MPOs. This program allowed to validate the inner channel Iridium coating (enhancing the reflectivity), the MPO slumping procedure, the Aluminium MPO surface coating, the thermal constraints for the optic (thermal control is needed) and the mechanical interfaces.
At the same time, detailed MPO examination allowed the Leicester team to improve the optical model of the MPOs in order to better reproduce the PSF imperfections due to channel deviations.  This improved knowledge of MPO characteristics will help to propose a better design for the THESEUS/SXI optics.

\subsubsection{Improving the MPOs TRL with the SVOM/MXT project} 
The MPOs are critical objects and are at the center of the THESEUS/SXI performances. In July 2016, the MXT optics bread board has been extensively tested in Leicester and at MPE Panter X-ray testing facility. The results of these tests allow us to affirm that the MPOs have already reached a TRL level of 4-5. During the summer of 2018, the MXT optics Structural and Thermal Model will be tested in Leicester and finally validated in Panter leading us to reach a TRL level of 6. End 2019 the MXT optics Flight Module will pass final performance validation in Panter. At this point the MPOs will have reached a TRL of 7. Then, the complete MXT telescope will be tested end 2020 in Panter leading us to a TRL level of 8 and finally end 2021-early 2022 MXT will be tested in flight (TRL 9). Thus at the time of the mission adoption (2021-2022) the Photonis MPOs will have largely reached the TRL level of 5 required for the critical elements.

\subsection{Demonstrator of the scientific performance of an NIR European detector} 
In the framework of the ALFA project (Astronomy Large Focal Plane Array), the SOFRADIR company in collaboration with ESA, CEA and the Focus Labex develops a flight quality 2k$\times$2k large-format NIR detector ``European Made''. This development supported by ESA could make it possible to have a European detector for the THESEUS IRT instrument (see G\"{o}tz et al., this proceedings).
By 2019 this development should reach a TRL 4 maturity. The first prototype will then be delivered to the SVOM consortium to equip the infrared channel of the SVOM robotic telescope installed in Mexico (F-GFT). In this context, from 2019, the first prototype of the ALFA development will be tested in detail on the sky. 
In parallel, the ALFA project will enter a new phase to spatialize the detector. This phase should make it possible to achieve a TRL of 6 in 2021 at the time of the THESEUS mission adoption.

\subsection{Reuse of the alert network developed for SVOM (45 stations)}
\begin{figure*}
\center
\includegraphics[width=\textwidth]{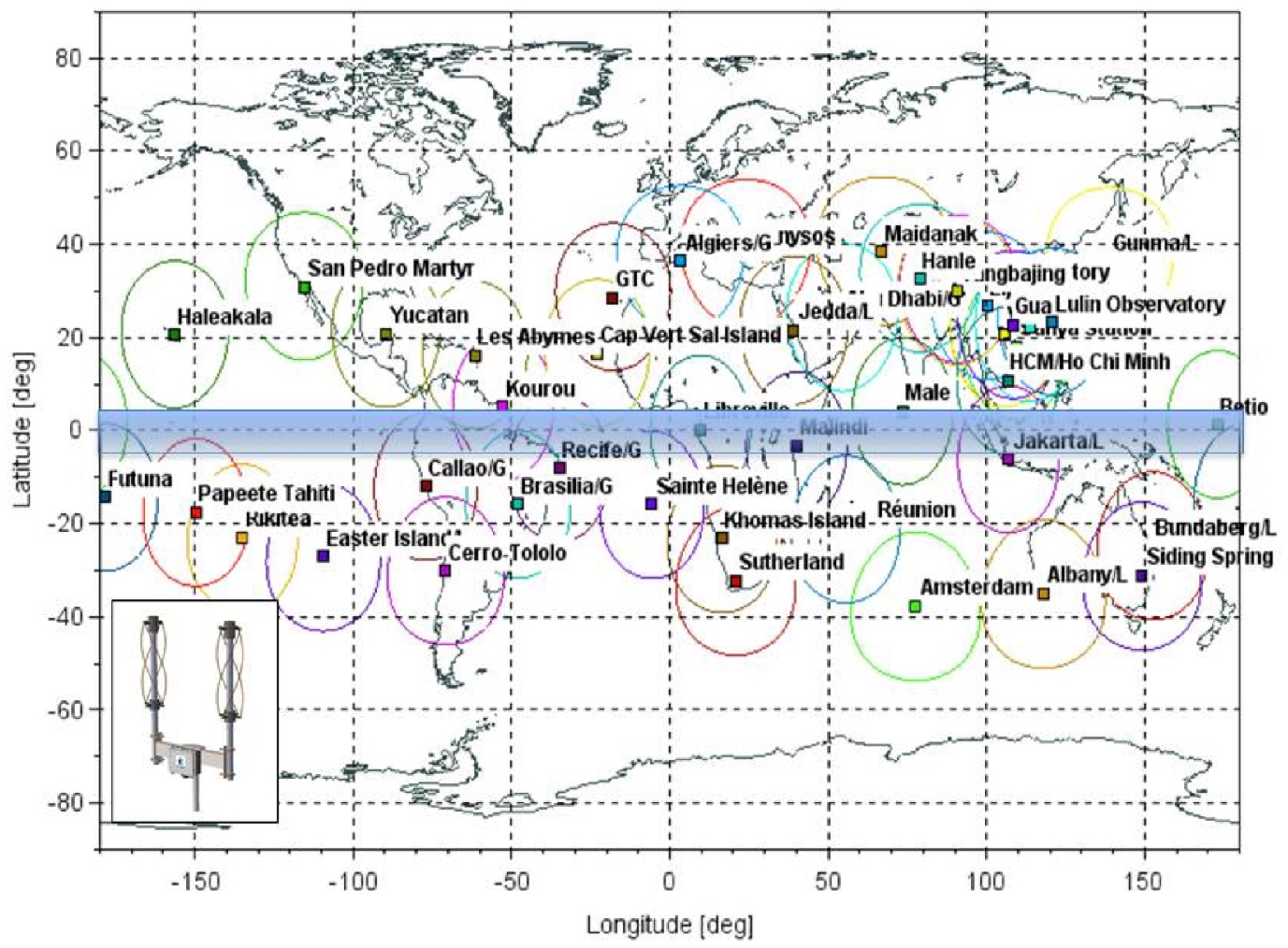}
{\caption{SVOM VHF Network. In blue the projected THESEUS orbit.\label{fig:VHF}}}
\end{figure*}

The SVOM alert network is composed of 45 VHF stations. This network will receive the alert from the satellite and boradcast them in less than 30 seconds from the on board localization time for 2/3 of the GRBs detected by ECLAIRs. The THESEUS mission will be able to reuse part of this this network. Due to the lower orbital inclination of THESEUS with respect to SVOM, the former will use only the low-latitude part of the network (approximately 20 stations, see Fig. \ref{fig:VHF}).

\section{Conclusions}

Following the path of the \swi, SVOM has been designed to be a highly versatile astronomy satellite, with built-in multi-wavelength capabilities, autonomous repointing and dedicated ground-based follow-up. Optimized observation and follow-up strategy is foreseen, aiming at redshift determination for a large fraction of SVOM GRBs ($>$50\%).
SVOM, and CTA/LSST/SKA, etc.. are promised to be great partners in the study of the transient electro-magnetic sky in the period 2022-2027. Thanks to its instrumental combination SVOM has the possibility to detect and localize a short GRB associated with a GW event, like the one detected in coincidence with GW170817. SVOM is hence prepared to play an important role in the multi-messenger era following-up GW, neutrino and VHE $\gamma$-ray alerts.
With a planned launch date end 2021 the SVOM mission will allow to develop and demonstrate the feasibility of several technical solutions 
that could be reused by the THESEUS mission. Those developments will be completed before the possible mission adoption of THESEUS by ESA.


\bibliography{SVOM_THESEUS.bib}
\bibliographystyle{aa}

\end{document}